\documentclass{ws-procs9x6}

\usepackage{epsfig}

\newcommand{\CK}{\v Cerenkov}

\begin{document}

\title{Isotope separation with the RICH detector of the AMS Experiment}

\author{L. Arruda, F. Bar\~ao, J. Borges, F. Carmo, P. Gon\c{c}alves,
  \underline{Rui Pereira}, M. Pimenta}
\address{LIP/IST \\
         Av. Elias Garcia, 14, 1$^o$ andar\\
         1000-149 Lisboa, Portugal \\
         e-mail: pereira@lip.pt}

\maketitle

\vspace{-0.7cm}

\abstracts{
The Alpha Magnetic Spectrometer (AMS), to be installed on the International Space
Station (ISS) in 2008, is a cosmic ray detector with several subsystems, one of which
is a proximity focusing Ring Imaging \CK\ (RICH) detector. This detector will be
equipped with a dual radiator (aerogel+NaF), a lateral conical mirror and a detection
plane made of 680 photomultipliers and light guides, enabling precise
measurements of particle electric charge and velocity. Combining velocity measurements
with data on particle rigidity from the AMS Tracker it is possible to obtain a
measurement for particle mass, allowing the separation of isotopes.
\\
A Monte Carlo simulation of the RICH detector, based on realistic properties
measured at ion beam tests, was performed to evaluate isotope separation capabilities.
Results for three elements --- H (Z=1), He (Z=2) and Be (Z=4) --- are presented.
}

\vspace{-0.9cm}

\section{The AMS02 experiment}

Alpha Magnetic Spectrometer (AMS)\cite{bib:ams} is an experiment designed to study the
cosmic ray flux by direct detection of particles above the Earth's atmosphere. The
deployment of the final detector (AMS-02) to the ISS is scheduled for 2008, for a minimum
operating period of 3 years. A preliminary version of the detector (AMS-01) was
successfully flown aboard the US space shuttle Discovery in June 1998.

On the ISS, orbiting at an average altitude of 400 km, AMS will collect an
extremely large number of cosmic ray particles. Its main goals are \emph{(i)} a detailed
study of cosmic ray composition and energy spectrum through the collection of an
unprecedented volume of data, \emph{(ii)} a search for heavy antinuclei (\mbox{$Z \geq$ 2})
which if discovered would signal the existence of antimatter domains in the Universe, and
\emph{(iii)} a search for dark matter constituents by examining possible signatures of
their presence in the cosmic ray spectrum.

\begin{figure}[htb]

\center

\vspace{-0.2cm}

\mbox{\epsfig{file=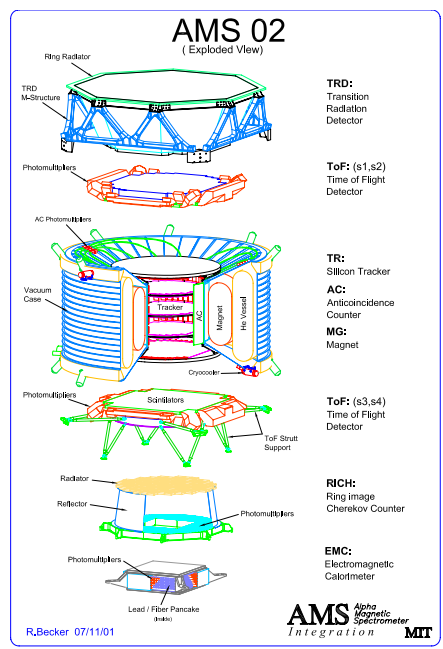,width=0.5\textwidth,clip=}}

\caption{Expanded view of the AMS-02 detector \label{amsdet}}

\vspace{-1.0cm}

\end{figure}

AMS is a spectrometer equipped with a superconducting magnet. It is composed of several
subdetectors: a Transition Radiation Detector (TRD), a Time-of-Flight (TOF) detector, a
Silicon Tracker, Anticoincidence Counters (ACC), a Ring Imaging \CK\ (RICH) detector and
an Electromagnetic Calorimeter (ECAL). Fig.~\ref{amsdet} shows a schematic view of the
full AMS detector. The present work evaluates reconstruction capabilities of the RICH
detector of AMS.

\section{RICH detector simulation}

The AMS RICH detector\cite{bib:buenerd} has a dual radiator configuration with a square
of sodium fluoride (NaF) with a refractive index $n$=1.334 at the centre surrounded by
tiles of silica aerogel with $n$=1.05. Detector efficiency is increased by the presence
of a highly reflective ($\approx$ 85\%) conical mirror surrounding the expansion
volume. For additional information on the RICH detector capabilities see also
ref.~\refcite{bib:luisa}.

A full-scale Monte Carlo simulation of the RICH detector was performed to evaluate
isotope separation capabilities using the GEANT 3 software package.
Data on particle rigidity, which in the experimental setup are expected to come from the
AMS silicon tracker, were created by adding a random smearing to the simulated rigidity.
The function giving smearing magnitude was adjusted to match real tracker performance.

Table \ref{simstat} shows the total number of events generated for each simulation.
The total number of simulated events corresponds to approximately one day of data, in the
cases of H and He, and one year in the case of Be. Simulated distributions were
based respectively on ref.~\refcite{bib:seo:h} for H, ref.~\refcite{bib:seo:he} for He and
ref.~\refcite{bib:strong:be} for Be and adjusted to the AMS detector acceptance.

%
%
%
%
%
%

\begin{table}[htb]

\vspace{-0.3cm}

\tbl{Statistics for the AMS RICH simulations}
{
\begin{tabular}{|r@{ : }l|r@{ : }l|r@{ : }l|}

\hline \multicolumn{6}{|c|}{\textbf{Simulations for NaF+aerogel}}

\\ \hline \hline \textbf{$\textrm{H}$ total} & $\mathbf{1.63 \times 10^7}$
& ${}^1 \textrm{H} \ (p)$ & $1.61 \times 10^7$
& ${}^2 \textrm{H} \ (d)$ & $1.39 \times 10^5$

\\ \hline \hline \textbf{$\textrm{He}$ total} & $\mathbf{2.02 \times 10^6}$
& ${}^3 \textrm{He}$ & $3.39 \times 10^5$
& ${}^4 \textrm{He}$ & $1.68 \times 10^6$

\\ \hline \hline \textbf{$\textrm{Be}$ total} & $\mathbf{8.47 \times 10^5}$
& ${}^9 \textrm{Be}$ & $6.97 \times 10^5$
& ${}^{10} \textrm{Be}$ & $1.49 \times 10^5$

\\ \hline \multicolumn{6}{c}{}

\vspace{-0.15cm}

\\ \hline \multicolumn{6}{|c|}{\textbf{Simulation for NaF only}}

\\ \hline \hline \textbf{$\textrm{H}$ total} & $\mathbf{1.53 \times 10^7}$
& ${}^1 \textrm{H} \ (p)$ & $1.52 \times 10^7$
& ${}^2 \textrm{H} \ (d)$ & $1.31 \times 10^5$
\\ \hline

\end{tabular}

\label{simstat}
}

\end{table}

\vspace{-0.2cm}

An additional simulation was performed for hydrogen events radiating in NaF, with a total
statistics corresponding to approximately one week of data taking. This was due to the
relatively low number of NaF events produced by generic simulations (NaF events correspond
to only $\sim$ 10\% of the total RICH data).

\section{Reconstruction procedure}

For each particle, charge and kinetic energy were determined using the procedure described in
ref.~\refcite{bib:barao2003}. Since isotopic ratios are a function of energy, the
reconstructed spectrum in energy-per-nucleon was divided in narrow regions for which the
calculation of isotopic abundances was performed separately. Only events with a minimum of 3
hits in the \CK\ pattern were considered. In the cases of He and Be, total isotopic abundances
for each energy bin were determined by fitting the mass spectrum to a sum of two gaussian
functions with an additional constraint on the mass resolutions
($\frac{\sigma_1}{\sigma_2} = \frac{m_1}{m_2}$).

Hydrogen was a special case due to low masses and the small $d/p$ ratio. An inverse mass ($1/m$)
spectrum was used, with separate fits being performed for the two mass peaks. A gaussian fit was
used for $p$, while for $d$ a sum of a gaussian and a constant was used to account for proton
background.

For all elements, final isotopic ratios were calculated from the gaussian fit integrals
corresponding to each peak.

\begin{figure}[htb]

\center

\vspace{-1.2cm}

\begin{tabular}{cc}  

\mbox{\epsfig{file=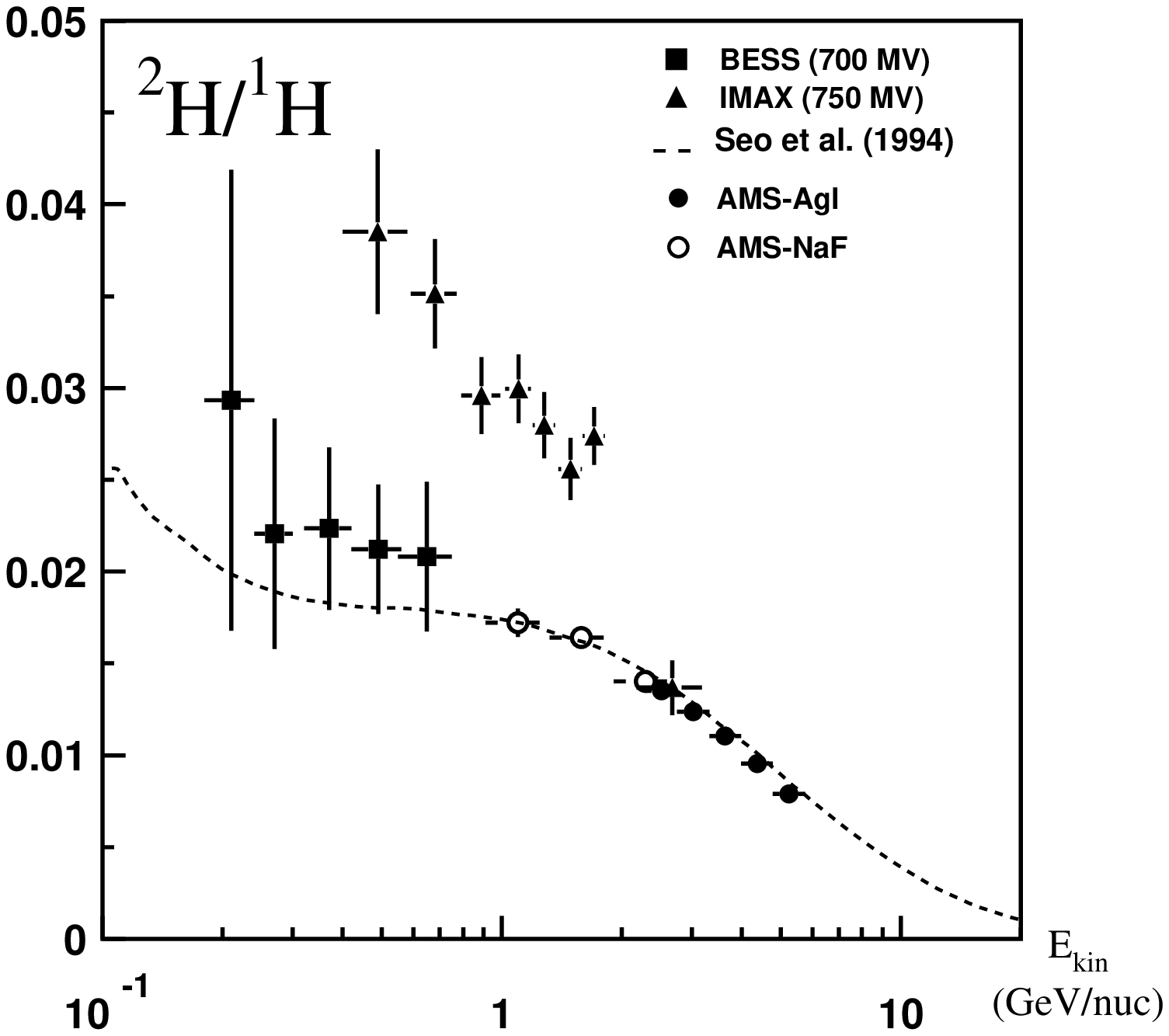,width=0.49\textwidth,clip=}}
&
\mbox{\epsfig{file=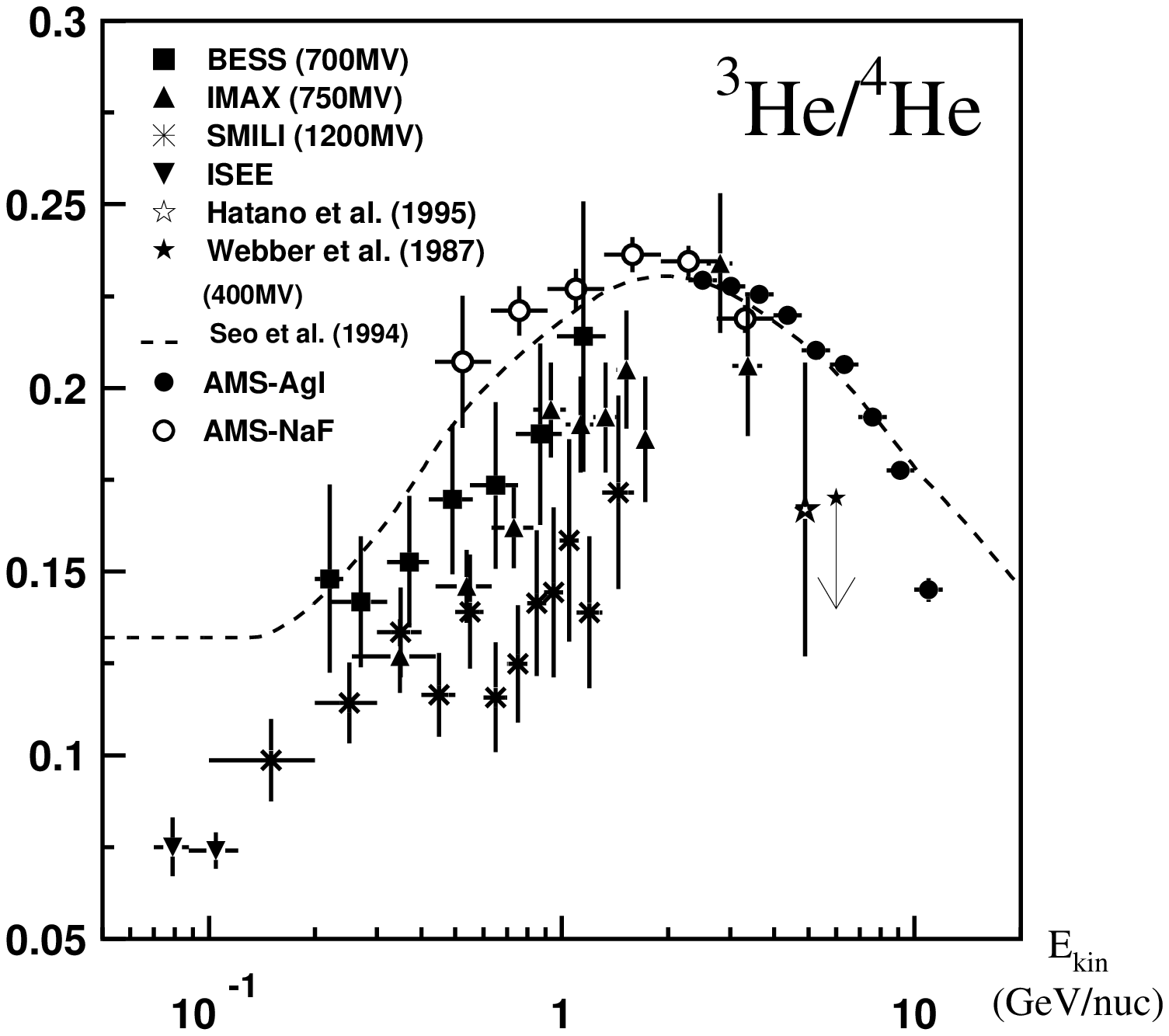,width=0.49\textwidth,clip=}}

\vspace{-1.0cm}
\\

\multicolumn{2}{c}{
\mbox{\epsfig{file=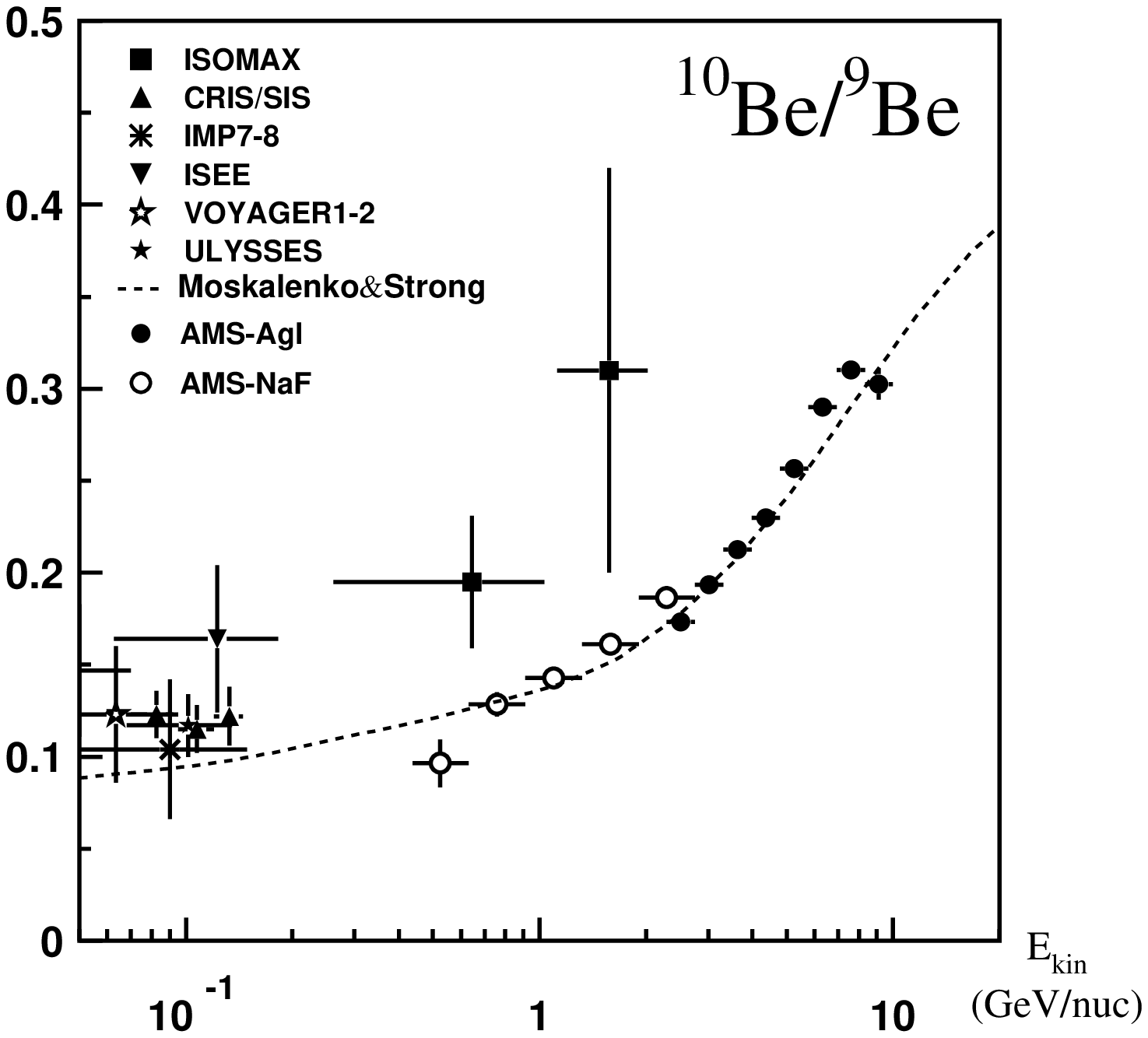,width=0.49\textwidth,clip=}}
}

\end{tabular}

\vspace{-0.5cm}

\caption{Reconstruction of simulated isotopic ratios in AMS
compared with data from other experiments \label{otherexp}}

\end{figure}

\section{Reconstruction results}

Fig.~\ref{otherexp} shows the results obtained for isotopic ratios compared with the
simulated distributions. Data from previous experiments are also shown for comparison. In the
cases of He and Be, satisfactory fits were obtained for the energy regions from the \CK\
thresholds up to $\sim$ 3 GeV/nucleon (NaF) and $\sim$ 10 GeV/nucleon (aerogel). In the case of H
good fits were
only obtained for the regions between 0.9 and 3 GeV/nucleon in NaF and from the \CK\ threshold
up to $\sim$ 6 GeV/nucleon in aerogel. These figures clearly show that even a small
fraction of the expected AMS statistics will represent a major improvement on existing results
for any of the three elements.

\begin{figure}[htb]

\center

\vspace{-0.7cm}

\begin{tabular}{cc}  

\mbox{\epsfig{file=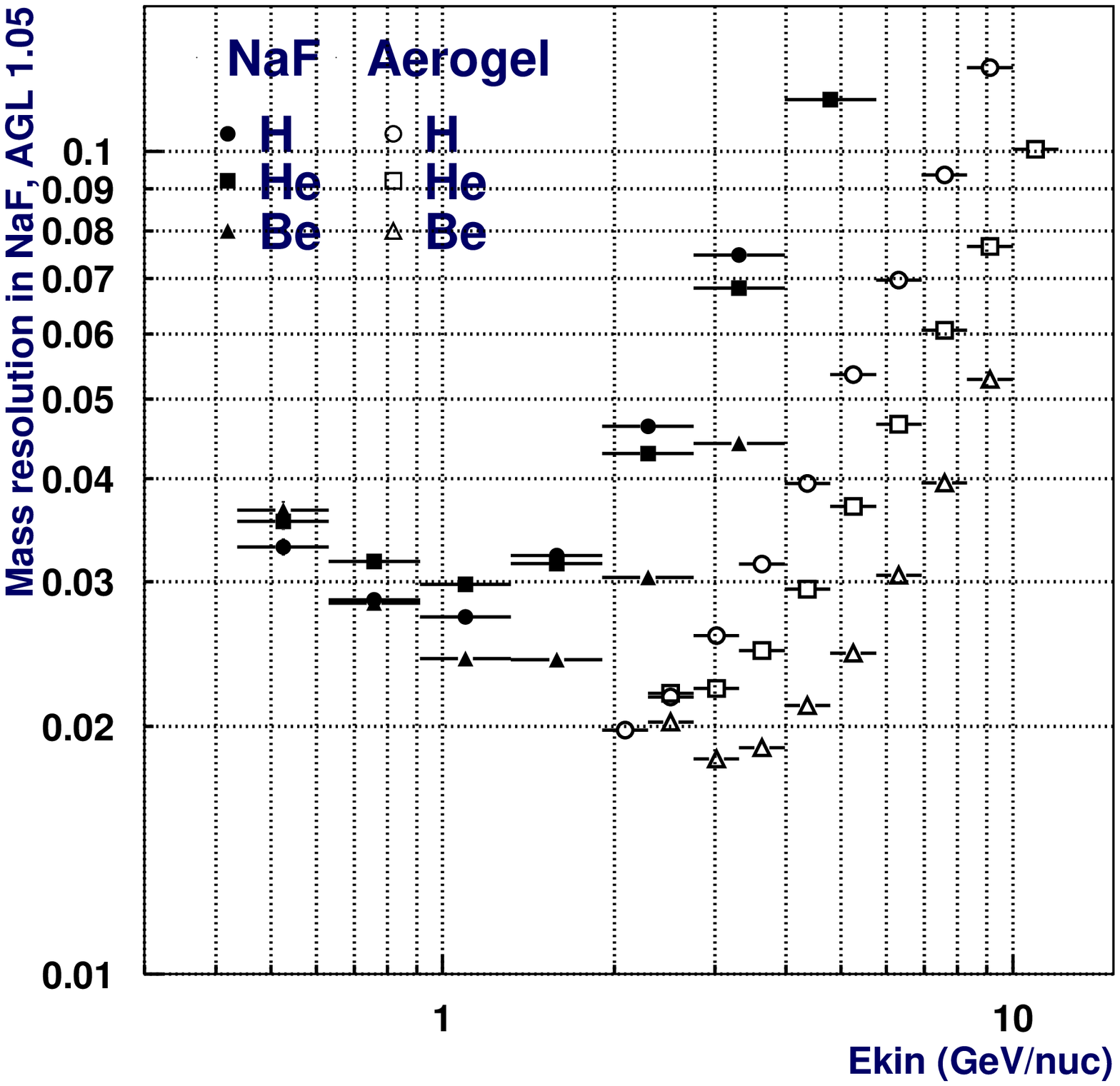,width=0.5\textwidth,clip=}}
&
\mbox{\epsfig{file=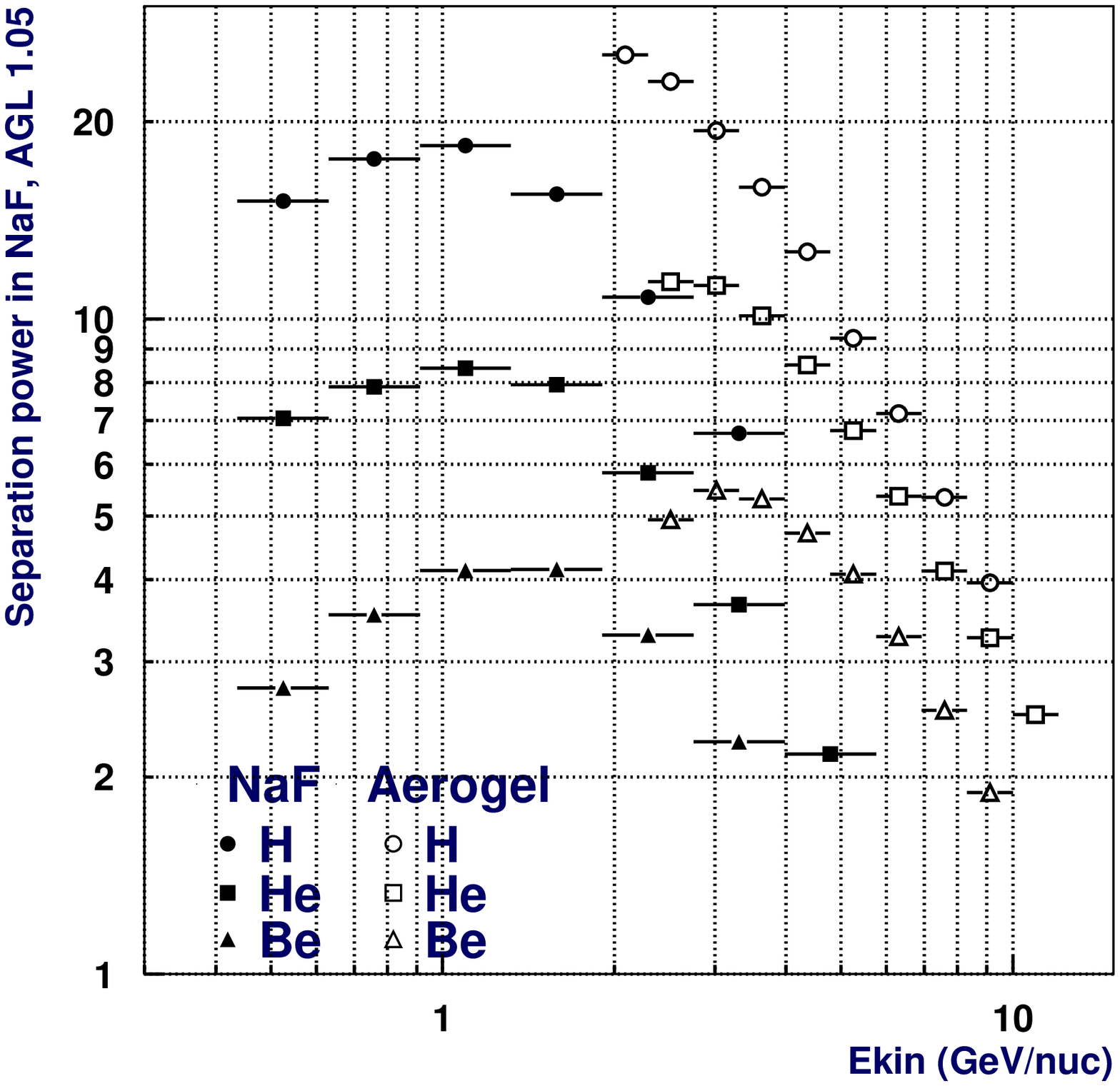,width=0.5\textwidth,clip=}}

\end{tabular}

\vspace{-0.2cm}

\caption{Simulation results for mass resolution \emph{(left)} and
separation power \emph{(right)}\label{mrsp}}

\vspace{-0.2cm}

\end{figure}

For each element, data on mass resolution and separation power for different energies were
obtained from fit results. Separation power was defined as the ratio $\frac{\Delta m}{\sigma_m}$.
Fig.~\ref{mrsp} shows mass resolution and separation power as functions of energy for both
radiators. Optimal mass resolutions were reached around 1 GeV/nucleon in NaF and 3 GeV/nucleon
in aerogel.

Separation power is higher for lighter elements, suggesting isotope separation should
be possible up to higher energies in the case of hydrogen. However, the greater difference
between proton and deuteron statistics ($d/p \sim 10^{-2}$) compared to the cases of He
and Be isotopes eventually leads to the separation being only possible up to $\sim$
6 GeV/nucleon compared to $\sim$ 10 GeV/nucleon for the other elements.

\vspace{-0.2cm}

\section{Conclusions}
AMS will provide a major improvement on existing data for isotopic abundances in
cosmic rays. Simulation results indicate that the separation of light isotopes using the
combination of RICH data and tracker rigidity measurements is feasible. The dual radiator
configuration of NaF and aerogel makes isotope separation of light elements possible for
energies in the range from 0.5 to 10 GeV/nucleon, approximately. Best mass resolutions
are $\sim$~2\% at 3 GeV/nucleon for aerogel, and $\sim$~3\% at 1 GeV/nucleon for NaF.
Techniques presented here may also be applied in the separation of antimatter isotopes
which is of great importance in dark matter studies.

\end{document}